\documentclass[aps,reprint,prl,amsmath,showpacs,showkeys,amssymb,amsfonts,
nofootinbib]{revtex4-1}

\usepackage{graphics}
\usepackage{xspace}
\DeclareMathAlphabet{\mathcal}{OMS}{cmsy}{m}{n}

\def\APJ{Astrophys.\ J.}
\def\APJL{Astrophys.\ J.\ Lett.}

\def\CQG{Classical Quant.\ Grav.}

\def\JCAP{J.\ Cosmol.\ Astropart.\ Phys.}

\def\JHEP{J.\ High Energ.\ Phys.}
\def\JMP{J.\ Math.\ Phys.}

\def\LRR{Living Rev.\ Relat.}

\def\MNRAS{Mon.\ Not.\ R.\ Astron.\ Soc.}%{M}on.\ {N}ot.\ {R}.\ {Astron}.\
\def\NAT{Nature}

\def\PRD{Phys.\ Rev.\ D}

\def\PRL{Phys.\ Rev.\ Lett.}
\graphicspath{{figures/}}
\usepackage{xspace}
\usepackage{amsmath}

\newcommand{\difftot}[2]{\frac{{\mathrm d}#1}{{\mathrm d}#2}}

\newcommand{\diffpar}[2]{\frac{{\partial}#1}{{\partial }#2}}

\newcommand{\e}[1]{{\mathrm e}^{#1}}

\begin{document}
\title{No-hair theorem for Black Holes in Astrophysical Environments}

\author{Norman G{\"u}rlebeck}
\affiliation{Center of Applied Space Technology and Microgravity (ZARM),
University of Bremen, Am Fallturm, 28359 Bremen, Germany, EU 
\email{norman.guerlebeck@zarm.uni-bremen.de}}


\begin{abstract}
According to the no-hair theorem, static
black holes are described by a Schwarzschild spacetime provided there are no
other sources of the gravitational field. This requirement, however, is in
astrophysical realistic scenarios often violated, e.g., if the black hole is
part of a binary system or if it is surrounded by an accretion disk. In
these cases, the black hole is distorted due to tidal forces. Nonetheless, the
subsequent formulation of the no-hair theorem holds: The
contribution of the distorted black hole to the multipole moments
that describe the gravitational field close to infinity and, thus, all
sources is that of a Schwarzschild black hole.
It still has no hair. This implies that there is no multipole moment induced in
the black hole and that its second Love numbers, which measure some aspects of
the distortion, vanish as was already shown in approximations to general
relativity. But here we prove this property for astrophysical relevant black
holes in full general relativity.
\end{abstract}
\maketitle

\section{Introduction}

The no-hair theorem states that any isolated static black hole is necessarily a
Schwarzschild black hole and that there is only one free parameter describing
the spacetime -- the mass $M$. Although the black hole has actually one hair,
$M$, this property is still called the no-hair theorem and, thus, the black hole is
called bald. This theorem is, of course, very appealing for astrophysics, since
just one parameter has to be measured to determine the entire spacetime around
a black hole. Even if the black hole rotates, a similar
theorem holds and only the mass and the spin have to be measured. 

In fact, these two parameters are already observable, see, e.g., Refs.\
\cite{Done_2013,Reynolds_2013,Risaliti_2013}. Additionally, new observatories
like GRAVITY (see Ref.\ \cite{Eisenhauer_2009}) will improve these measurements
further. An independent and promising approach is the measurement of the shadow
of a black hole, see, e.g., Refs.\ \cite{Grenzebach_2014,Li_2014} for recent
results. The shadow will be resolvable in the millimeter/submillimeter range
with the Event Horizon Telescope \cite{Doeleman_2008}. Moreover, the potential
discovery of a binary system containing a black hole promises headway for
characterizing black holes either via pulsar timing if the companion is a pulsar
or via the detection of gravitational waves.

Since the no-hair theorem dictates in the rotating case that the quadrupole
moment of a black hole is determined by its mass and spin, an independent
measurement of all three parameters allows for a test of alternative theories of
gravity or of the assumptions of the no-hair theorem, for recent approaches see,
e.g., Refs.\ \cite{Broderick_2013,Johannsen_2013,Wade_2013}. For example, the source
could be described by matter distributions like boson stars for which
the no-hair theorem does not hold. 

A crucial assumption for the
no-hair theorem is that the black hole is isolated, i.e., the spacetime is
asymptotically flat and contains no other sources. However, in many
astrophysical situations this requirement is not fulfilled, e.g. for black holes
in binary systems, if the black hole is surrounded by plasma, an accretion disk
or if jets are formed in its vicinity, i.e., it might put on different types of
wigs. These additional sources contribute also to the total multipole moments of the spacetime. Hence, a formulation like the standard no-hair theorem for isolated black holes cannot hold anymore, whereas a formulation solely for the \emph{part} of the total multipole moment sourced by the black hole might still be correct. The latter is shown in this Letter.
In the exterior field of additional sources, the black hole is distorted and the
inner geometry of the horizon changes. This is measured by the Love numbers of
the first kind or the multipole moments of isolated horizons, see Refs.\ 
\cite{Ashtekar_2004,Ashtekar_2004a,Damour_2009a}.
Nonetheless, we show here that this does not imply that the black holes are not
anymore bald. More precisely: Although the total multipole moments of the
spacetime measured at infinity change, this is solely due to the external
sources and not to a different contribution of the black holes themselves. In fact, \emph{distorted black holes have only a mass monopole}. Thus, even though the black hole might put on a wig it still looks bald. 
Note that we assume here for simplicity static black holes.

Distortions of static black holes and neutron stars are of particular interest
for inspirals treated in an adiabatic regime, see, for details on the
validity of this regime, Refs.\ \cite{Flanagan_2008,Damour_2009}.
In such a quasistatic approximation, the black hole or the neutron star is
distorted due to the external field of the companion instantaneously. Thus, the
system is additionally axially symmetric with respect to the axis joining
the two constituents of the binary system and the metric of distorted black
holes of Ref.\ \cite{Geroch_1982} is applicable.
The imprints of such distortions in the gravitational waves emitted by
inspiraling binaries give information on the equation of state of neutron stars
\cite{Flanagan_2008,Hinderer_2008}. Moreover, those imprints
can be used to experimentally reveal if a constituent of a binary system is a black
hole. This provides an avenue along which the existence of black holes can be
directly inferred. On the other hand, if
the existence of a black hole in a binary system is established
independently by observing, say, gamma ray bursts at later stages of the inspiral, then the measurement of
its distortions using gravitational waves allows us to test general
relativity via the here presented no-hair theorem.

The distortions of the black holes and neutron stars are characterized by the
Love numbers of first and second kind, $h_r$ and $k_r$, cf.\ Refs.\
\cite{Love_1911,Binnington_2009,Damour_2009}.
Roughly speaking, the $h_r$ measure the changes in the shape of the horizon
or the neutron star and the $k_r$ measure the change in the asymptotic
multipole moments caused by the distortion due to an external source, see Refs.\
\cite{Binnington_2009,Damour_2009a,Damour_2009,Kol_2012,Chakrabarti_2013}
for their use in general relativity. In the latter four works, it was
established using approximation methods that the $k_r$ vanish for four-dimensional black holes.
However, it was debated if this result is still valid in the case higher orders
in the approximations scheme are taken into account. We resolve
this dispute here by proving the result analytically without any approximation.
Although these distortions are not crucial to detect gravitational waves in prospective data from Advanced
LIGO and Advanced Virgo \cite{Berry_2014}, they will be important for a detailed analysis of the data and for future detectors with an increased sensitivity, see, e.g., Refs.\ \cite{Baiotti_2010,Read_2013}.

The Love numbers of the
second kind were also applied to establish universal relations, i.e., relations
that are independent of the equation of state, between certain physical
parameters describing neutron stars, see Refs.\
\cite{Yagi_2013,Yagi_2013a,Maselli_2013} but also Refs.\
\cite{Doneva_2014,Haskell_2014,Pappas_2014,Chakrabarti_2014}.
The here considered black hole case is solved analytically in full general relativity and, thus, it serves
as a test for the various approximation schemes employed for neutron stars also
in this respect.

Subsequently, we use geometric units, in which $G=c=1$, where $c$ is the velocity of
light and $G$ Newton's gravitational constant. The metric has the signature
$(-1,1,1,1)$. Greek indices run from $0$ to $3$ and Latin indices run from $1$
to $3$.

\section{Distorted Black Holes}\label{sec:distorted_black_hole}

The metric of arbitrary static and
axially symmetric spacetimes can be written in the Weyl form
under standard assumptions, cf.\ Ref.\ \cite{Stephani_2003}:
\begin{align}\label{eq:Weyl_line_element}
\begin{split}
  ds^2=\mathrm{e}^{2 k-2U}\left(d\rho^2+d\zeta^2\right)+
  W^2\mathrm{e}^{-2U} d\varphi^2-\mathrm{e}^{2U}d t^2,
\end{split}
\end{align}
where the functions $U,~k$, and $W$ depend on $\rho$ and $\zeta$. Note that the
metric functions $U$ and $W$ can be expressed by the timelike Killing vector
$\xi^\alpha$ and the spacelike Killing vector $\eta^\alpha$:
\begin{align}\label{eq:Killing_Vectors}
\begin{split}
  \mathrm{e}^{2U}=-\xi_\alpha\xi^\alpha,\quad W^2=-\eta_\alpha\eta^\alpha
  \xi_\beta\xi^\beta.
\end{split}
\end{align}

In case the exterior sources are static and axially symmetric or allow for a
quasistatic description, the general metric near the horizon $\mathcal H$ of a
distorted black hole was found by Geroch and Hartle in Ref.\ \cite{Geroch_1982} in
the form of Eq.\ \eqref{eq:Weyl_line_element}. In a neighborhood of $\mathcal H$, we
assume pure vacuum, which is physically reasonable if the matter can be treated 
quasistatically and satisfies the energy conditions, cf.\ Ref.\ \cite{Bardeen_1973}.
Thus, there exists a surface $\mathcal S_{\mathcal H}$, which encloses $\mathcal
H$ and no other sources. If $\mathcal S_{\mathcal H}$ is sufficiently close to
$\mathcal H$, the metric functions in Eq.\ \eqref{eq:Weyl_line_element} read
between $\mathcal S_{\mathcal H}$ and $\mathcal H$
\begin{align}\label{eq:Distorted_black_hole_line_element}
\begin{split}
  U=U_S+U_D,\quad k=k_S+k_{SD},\quad W=\rho.
\end{split}
\end{align}
The functions $U_S$ and $k_S$ are given by the respective quantities of the
Schwarzschild black hole:
\begin{align}\label{eq:Schwarzschild}
\begin{split}
  U_S&=\frac{1}{2}\log\left[\frac{r_{+}+r_{-}-2M}{r_{+}+r_{-}+2M}\right],\\
  k_S&=\frac{1}{2}\log\left[\frac{(r_{+}+r_{-})^2-4M^2}{4r_{+}r_{-}}\right],\\
  r_{\pm}^2&=\rho^2+(\zeta\pm M)^2.
\end{split}
\end{align}
Subsequently, we find that the parameter $M$ coincides with the
Komar mass of the distorted black hole. The function $U_D$ is determined by the
exterior matter and it solves a Laplace equation
\begin{align}\label{eq:Laplace}
\begin{split}
  \left(\diffpar{{}^2}{\rho^2}+\frac{1}{\rho}\diffpar{}{\rho}+
  \diffpar{{}^2}{\zeta^2}\right)U_D=0.
\end{split}
\end{align}
If $U_D$ vanishes, the spacetime describes a Schwarzschild black hole. The
function $k_{SD}$ follows from a line integration once $U_D+U_S$ is known,
cf.\ Ref.\ \cite{Stephani_2003}. However, we do not require its explicit form
subsequently. 

The horizon of the distorted black hole is located at the symmetry axis
($\rho=0$, $\zeta\in \left[-M,M \right]$) like for the Schwarzschild black hole.
In fact, in canonical Weyl coordinates the horizon can always be located at
$\rho=0$, see Ref.\ \cite{Carter_1973}. 
These coordinates allow a shift in the $\zeta$ coordinate. We
employed this freedom to place the horizon symmetrically with respect to that
coordinate, i.e., that the ``north or south pole'' of the horizon are characterized
by $\zeta_{N/S}=\pm M$. At these points, $U_D$
has to take the same value to avoid struts, which we want to exclude for simplicity, see
Ref.\ \cite{Geroch_1982}. If the
external matter is reflection symmetric, like for accretion disks or jets, this
is trivially satisfied. Note that the metric functions take the form of Eq.\
\eqref{eq:Distorted_black_hole_line_element}-\eqref{eq:Laplace} only
in a neighborhood of $\mathcal H$ and they neither describe directly the
asymptotic behavior nor the metric in the interior of the external source.
Nonetheless, we will be able to conclude with the help of the source integrals
the contributions of the distorted black hole to the asymptotic multipole
moments without specifying the exterior sources in detail. These could also
include other black holes. We only require that the spacetime is asymptotically
flat and that all \emph{external} sources are contained in a region, which does not contain $\mathcal H$ and which does not extend to infinity. We denote its boundary by $\mathcal S_{\mathrm{ext}}$.  

\section{The source integrals}

To disentangle the contributions of the black hole and the external sources to
the asymptotic multipole moments, the source integrals proved to be the
essential tool. They were recently derived in Ref.\ \cite{Gurlebeck_2013} and they
make it possible to define the asymptotics of the spacetime including the Geroch
multipole moments by evaluating quasilocal surface or volume integrals. The
respective surfaces and volumes need only to envelope or contain all regions
with a nonvanishing stress-energy tensor. Here we need the
surface integrals and introduce the required quantities, subsequently.

The Weyl multipole moments $U^{(r)}$ are defined as the expansion of $U$ along
the axis of symmetry close to infinity, i.e.,
\begin{align}\label{eq:Weyl_multipoles}
U=\sum\limits_{r=0}^\infty \frac{U^{(r)}}{\left|\zeta\right|^{r+1}}.
\end{align}
As we will see later the coordinate $\zeta$ can be defined geometrically so that
this definition is also covariant. Indeed, it was shown in Ref.\ \cite{Fodor_1989}
that from the $U^{(r)}$ the Geroch multipole moments $m_r$ can be determined
uniquely by nonlinear algebraic relations. To calculate the
$m_r$, the $U^{(k)}$ need to be known for $0\leq k\leq r$. Thus, it is sufficient for us to consider here the $U^{(r)}$. Note that the origin with respect to which the multipole moments are measured was chosen by requiring $\zeta_{N/S}=\pm M$.  

Furthermore, we use the functions
\begin{align}\label{eq:expansion_coefficients}
\begin{split}
  N_{-}^{(r)}(x,y)&=
  \sum\limits_{k=0}^{\left\lfloor\frac{r}{2}\right\rfloor}\frac{2(-1)^{k+1}
  r!x^{2k+1}y^{r-2k}}{4^k (k!)^2(r-2k)!},\\
  N_{+}^{(r)}(x,y)&=
  \sum\limits_{k=0}^{\left\lfloor\frac{r-1}{2}\right\rfloor}\frac{2
  (-1)^{k+1}r!x^{2k+2}y^{r-2k-1}}{4^k (k!)^2(r-2k-1)!(2k+2)}.
\end{split}
\end{align}
It can easily be checked that these functions
obey the equations
\begin{align}\label{eq:differential_equations_N}
\begin{split}
 N^{(r)}_{+,x}-N^{(r)}_{-,y}&=0,\quad
 N^{(r)}_{+,y}+N^{(r)}_{-,x}-\frac{N^{(r)}_{-}}{x}=0.
\end{split}
\end{align}
Commas denote partial derivatives. Additionally, let us introduce the 1-form
\begin{align}\label{eq:definition_Z}
  Z_\alpha=\epsilon_{\alpha\beta\gamma\delta}W^{,\beta}W^{-1}
  \eta^\gamma\xi^\delta,
\end{align}
where $\epsilon_{\alpha\beta\gamma\delta}$ is the volume form of the spacetime.
In vacuum, $Z_\alpha$ is exact and it is hypersurface orthogonal in the entire
spacetime. Since the surfaces of interest, $\mathcal S_{\mathcal H}$ and
$\mathcal S_{\mathrm{ext}}$, lie in the vacuum region or its boundaries, we can
introduce a scalar $Z$ via $Z_{,\alpha}=Z_{\alpha}$, for technical details and
a more general treatment see Ref.\ \cite{Gurlebeck_2013}. It turns out that $Z=\zeta$
in canonical Weyl coordinates if the constant of integration is suitably
chosen.

With this notation at hand, we can express the Weyl multipole moments by
\begin{align}\label{eq:surface_integrals}
\begin{split}
 U^{(r)}=&\int\limits_{\mathcal S_{\mathcal H}} \eta^{(r)}_a\hat n^a
  \mathrm d {\mathcal S}_{\mathcal H}+\int\limits_{\mathcal S_{\mathrm{ext}}}
  \eta^{(r)}_a\hat n^a \mathrm d {\mathcal S}_{\mathrm{ext}},\\
\eta^{(r)}_a=&\frac{1}{8\pi}\frac{\e{U}}{W}
\left(N_{-}^{(r)}U_{,a}-N^{(r)}_{+,W}Z_{,a}U+N^{(r)}_{+,Z}W_{,a}U\right),
\end{split}
\end{align}
where $\hat n^a$ denotes the outward pointing unit normal to the surfaces
$\mathcal S_{\mathcal H}$ and $\mathcal S_{\mathrm{ext}}$ and the functions
$N^{(r)}_\pm$ depend on $(x,y)=(W,Z)$, see \cite{Gurlebeck_2013}. $\mathrm d
{\mathcal S}_{\mathcal H}$ and $\mathrm d \mathcal S_{\mathrm{ext}}$ are the 
proper area elements of ${\mathcal S}_{\mathcal H}$ and $\mathcal
S_{\mathrm{ext}}$, respectively. In vacuum, we can always choose canonical Weyl
coordinates such that $W=\rho$ and $Z=\zeta$.

\section{The Induced Multipole Moments of Distorted Black Holes}

With Eq.\ \eqref{eq:surface_integrals}, we can identify the
contribution of the different sources to the asymptotic Weyl multipole moments
covariantly. The first term in Eq.\ \eqref{eq:surface_integrals}, which we denote
$U^{(r)}_{\mathcal H}$, gives the contribution of the distorted black hole and
the second term, $U^{(r)}_{\mathrm{ext}}$, the contribution of the external
sources. The induced multipole moment of a distorted black hole is now simply
defined as $U_{\mathrm{ind}}^{(r)}=U^{(r)}_{\mathcal H}-U^{(r)}_{S}$, where the
$U^{(r)}_{S}$ are the Weyl multipole moments of an undistorted Schwarzschild
black hole. They coincide with the Newtonian multipole moments of a line mass of
uniform density, see Ref.\ \cite{Stephani_2003}. We parameterize $\mathcal
S_{\mathcal H}$ for constant angles $\varphi$ from the ``north pole'' to the
``south pole" ($ s \in [ s_N, s_S]\mapsto
(\rho(s),\zeta(s),\varphi=\mathrm{const.}))$, cf.\ Ref.\ \cite{Gurlebeck_2013}.
Then we obtain with Eq.\ \eqref{eq:Distorted_black_hole_line_element} the
$U^{(r)}_{\mathcal H}$:
\begin{align}\label{eq:surface_integral}
\begin{split}
  U^{(r)}_{\mathcal H}&=\frac{1}{4}\int\limits_{s_N}^{s_S}
   \left[N^{(r)}_{-}\left(U_{S}+U_{D}\right)_{,n}-\right.\\
   &\left.\left(N^{(r)}_{+,W}Z_{,
   n}-N^{(r)}_{+,Z}W_{,
   n}\right)\left(U_{S}+U_{D}\right)\right] \mathrm d s,
\end{split}  
\end{align}
where we denote by $f_{,n}$ the normal derivative
$-f_{,\rho}\difftot{}{s}\zeta(s)+f_{,\zeta}\difftot{}{s}\rho(s)$. The multipole
moments of a Schwarzschild black hole can be inferred from Eq.\
\eqref{eq:surface_integral} by setting $U_D=0$. For the
induced multipole moments, we have in turn only to subtract this Schwarzschild
contribution from Eq.\ \eqref{eq:surface_integral} and get
\begin{align}
\begin{split}
  U_{\mathrm{ind}}^{(r)}&=\frac{1}{4}\int\limits_{s_N}^{s_S}
   \left[N^{(r)}_{-}U_{D,n}-N^{(r)}_{+,W}Z_{,n}U_{D}+\right.\\
   &\left. N^{(r)}_{+,Z}W_{,n}U_{D}\right] \mathrm d s.
\end{split}
\end{align}
Applying the divergence theorem and Eq.\ \eqref{eq:Laplace}, we can rewrite
$U^{(r)}_{\mathrm{ind}}$:
\begin{align}
\begin{split}
U^{(r)}_{\mathrm{ind}}
&=\frac{1}{8\pi}\int\limits_{V_{\mathcal
H}}\frac{1}{\rho}\left[U_{D,\rho}\left(N_{-,\rho}^{(r)}+
N_{+,\zeta}^{(r)}-\frac{N_{-}^{(r)}}{\rho}\right) + \right.\\ 
&\left. U_{D,\zeta}\left(N_{-,\zeta}^{(r)}-N_{+,\rho}^{(r)}\right)\right]\mathrm
d V_{\mathcal H},
\end{split}
\end{align}
which vanishes by virtue of Eq.\ \eqref{eq:differential_equations_N}.
$V_{\mathcal H}$ is the coordinate volume enclosed by $\mathcal S_{\mathcal H}$
and $\mathcal H$ in canonical Weyl coordinates. Thus, \emph{the induced
multipole moments vanish and the contribution of the distorted black hole to the
asymptotic Weyl multipole moments is the same as that of a Schwarzschild black
hole}. With the results in Refs.\ \cite{Fodor_1989, Gurlebeck_2013}, this can readily
be translated to Geroch's multipole moments. This generalizes the no-hair theorem
to black holes that are distorted by external matter. Note that this matter sources the gravitational field, too. Hence, the total asymptotic multipole moments differ in general from those of a Schwarzschild spacetime. This holds in particular for systems of two black holes. In the above derivation, the origin with respect to which the multipole moments are measured is chosen in the center of one of the black holes. Thus, the other black hole regarded as external matter contributes, for instance, a nonvanishing quadrupole moment to the total one, cf.\ the comment after Eq.\ \eqref{eq:Weyl_multipoles}.

The vanishing of the induced multipole moments implies that the second Love
numbers $k_r$ vanish, too, because they are proportional to
$U^{(r)}_{\mathrm{ind}}$. This is corroborated by the results in Refs.\
\cite{Damour_2009,Binnington_2009,Chakrabarti_2013,Kol_2012}.
But here we did not use any approximation or linearization. The result holds in
\emph{full general relativity}. Thus, one can assume $k_r=0$ for black holes, which rotate sufficiently slowly, in binary systems when calculating the emitted gravitational radiation during the adiabatic regime. Note that $k_r=0$ is specific to black holes. It
does not hold for neutron stars, cf.\ Ref.\ \cite{Binnington_2009}. Nonetheless, the
source integrals of the Weyl multipole moments are still tailored to calculate
their $k_r$, since the contributions from the individual sources to the Weyl
multipole moments are separated covariantly and a definition of an induced
multipole moment becomes possible in full general relativity.
 Moreover, it simplifies the evaluation of the
source integrals in Ref.\ \cite{Gurlebeck_2013} in the presence of black holes
considerably, since we have only to calculate the mass of the individual black holes to know all $U^{(r)}_{\mathcal H}$.

If the black holes rotate sufficiently slowly, there are several implications for astrophysics:
On the one hand, measuring the mass of the black hole determines its contribution to the
multipole moments completely. In binary systems containing a black hole or
for a black hole with an accretion disc, the mass of the black hole can be
inferred from the mass of the entire system measured by the motion of distant
stars and the mass of the companion star or disc. After that all multipole
moments of the black hole are fixed. Thus, every measurement of the multipole
moments of the entire system, say, the quadrupole moment determines the quadrupole moment of the companion or the disc.
 On the other hand, if the quadrupole moment of both, the entire system and the companion star or disc can be
measured, then general relativity can be tested. 

\section{The multipole moments of the horizon}

Whereas the distorted black hole has the same asymptotic multipole moments as
a Schwarzschild black hole, the horizon geometry clearly changes. This can be
easily seen by evaluating the covariantly defined multipoles $M_{n}$ of isolated
horizons following Ref.\ \cite{Ashtekar_2004,Ashtekar_2004a}.
The scheme outlined therein was independently carried out in Ref.\ \cite{Damour_2009a}
using Schwarzschild-like coordinates. In that paper, it was found that the
multipole moments of the distorted horizon are different from those of the
Schwarzschild black hole. In fact, these deviations were used to define a01
relativistic analogue of the first Love numbers for black holes, which do not
vanish in contrast to the second Love numbers.

The change in the geometry of the horizon is, however, not reflected in the
asymptotic multipole moments. This, at first glance, counterintuitive behavior
can be understood with a trivial Newtonian example.
Consider a point mass and its multipoles. All multipole moments but the mass
vanish and the equipotential surfaces are spheres.
If an additional gravitational field generated by, say, a second point mass
separated from the first is introduced, the multipole moments of the original
point mass, which can be evaluated with Newtonian source integrals, are
unchanged.
In fact, the point particle has no inner structure and, thus, cannot be
distorted by an external gravitational field and the source stays the same.
Nonetheless, the equipotential surfaces are no longer spheres analogously to the
distorted horizon.
The situation changes of course, if an internal structure is assumed, like one
would have to do for the description of neutron stars. Then the external
gravitational field can indeed deform the matter distribution and the sources,
which can be measured in the asymptotic multipole moments.

\begin{acknowledgments}
N.G. gratefully acknowledges support from the DFG within the Research
Training Group 1620 ``Models of Gravity''. Partial support comes also from "NewCompStar", COST Action MP1304. The author thanks A. Ashtekar and J. Steinhoff for helpful discussions.
\end{acknowledgments}

%\bibliographystyle{h-physrev}%plaindin
%\bibliography{References}

\begin{thebibliography}{10}

\bibitem{Done_2013}
C.~Done, C.~Jin, M.~Middleton, and M.~Ward,
\newblock \MNRAS {\bf~434}, 1955 (2013).

\bibitem{Reynolds_2013}
C.~S. {Reynolds},
\newblock \CQG {\bf~30}, 244004 (2013).

\bibitem{Risaliti_2013}
G.~{Risaliti} {\em et~al.},
\newblock \NAT {\bf~494}, 449 (2013).

\bibitem{Eisenhauer_2009}
F.~{Eisenhauer} {\em et~al.},
\newblock {GRAVITY: Microarcsecond Astrometry and Deep Interferometric Imaging
 with the VLT},
\newblock in {\em Science with the VLT in the ELT Era}, edited by
  A.~{Moorwood}, Springer, Netherlands, p. 361, 2009.

\bibitem{Li_2014}
Z.~{Li} and C.~{Bambi},
\newblock \JCAP {\bf~1}, 41 (2014).

\bibitem{Grenzebach_2014}
A.~{Grenzebach}, V.~{Perlick}, and C.~{L\"ammerzahl},
\newblock \PRD {\bf 89}, 124004 (2014).

\bibitem{Doeleman_2008}
S.~S. {Doeleman} {\em et~al.},
\newblock \NAT {\bf~455}, 78 (2008).

\bibitem{Broderick_2013}
A.~E. Broderick, T.~Johannsen, A.~Loeb, and D.~Psaltis,
\newblock \apj {\bf~784}, 7 (2014).

\bibitem{Johannsen_2013}
T.~{Johannsen} and D.~{Psaltis},
\newblock \APJ {\bf~773}, 57 (2013).

\bibitem{Wade_2013}
M.~{Wade}, J.~D.~E. {Creighton}, E.~{Ochsner}, and A.~B. {Nielsen},
\newblock \PRD {\bf~88}, 083002 (2013).

\bibitem{Ashtekar_2004}
A.~Ashtekar, J.~Engle, T.~Pawlowski, and C.~v.~d. Broeck,
\newblock \CQG {\bf~21}, 2549 (2004).

\bibitem{Ashtekar_2004a}
A.~Ashtekar and B.~Krishnan,
\newblock \LRR {\bf~7}, (2004).

\bibitem{Damour_2009a}
T.~{Damour} and O.~M. {Lecian},
\newblock \PRD {\bf~80}, 044017 (2009).

\bibitem{Flanagan_2008}
E.~E. Flanagan and T.~Hinderer,
\newblock \PRD {\bf~77}, 021502(R) (2008).

\bibitem{Damour_2009}
T.~{Damour} and A.~{Nagar},
\newblock \PRD {\bf~80}, 084035 (2009).

\bibitem{Geroch_1982}
R.~{Geroch} and J.~B. {Hartle},
\newblock \JMP {\bf~23}, 680 (1982).

\bibitem{Hinderer_2008}
T.~Hinderer,
\newblock \APJ {\bf~677}, 1216 (2008),
\newblock \APJ~{\bf 697}, 964(E) (2009).

\bibitem{Love_1911}
A.~Love,
\newblock {\em Some Problems of Geodynamics} (Cornell University Library,
  Ithaca, 1911).

\bibitem{Binnington_2009}
T.~{Binnington} and E.~{Poisson},
\newblock \PRD {\bf~80}, 084018 (2009).

\bibitem{Kol_2012}
B.~{Kol} and M.~{Smolkin},
\newblock \JHEP {\bf~2}, 10 (2012).

\bibitem{Chakrabarti_2013}
S.~{Chakrabarti}, T.~{Delsate}, and J.~{Steinhoff},
\newblock arXiv:1304.2228 [gr-qc].

\bibitem{Berry_2014}
C.~P.~L. {Berry}, I. {Mandel}, H. {Middleton}, L.~P. {Singer},  A.~L. {Urban}, A. {Vecchio}, S. {Vitale}, K. {Cannon}, B. {Farr}, W.~M. {Farr}, P.~B. {Graff}, C. {Hanna}, C.-J. {Haster}, S. {Mohapatra}, C. {Pankow}, L.~R. {Price}, T. {Sidery}, and J. {Veitch},
\newblock arXiv:1411.6934 [astro-ph].


\bibitem{Baiotti_2010}
L.~{Baiotti}, T.~{Damour}, B.~{Giacomazzo}, A.~{Nagar}, and L.~{Rezzolla},
\newblock \PRL {\bf 105}, 261101 (2010).

\bibitem{Read_2013} 
J.~S.~{Read}, L.~{Baiotti}, J.~D.~E.~{Creighton}, J.~L.~{Friedman},
B.~{Giacomazzo},  K.~{Kyutoku}, C.~{Markakis}, L.~{Rezzolla}, M.~{Shibata},
and K.~{Taniguchi}
\newblock \PRD {\bf 88}, 044042 (2013).

\bibitem{Yagi_2013}
K.~Yagi and N.~Yunes,
\newblock Science {\bf~341}, 365 (2013).

\bibitem{Yagi_2013a}
K.~{Yagi} and N.~{Yunes},
\newblock \PRD {\bf~88}, 023009 (2013).

\bibitem{Maselli_2013}
A.~{Maselli}, V.~{Cardoso}, V.~{Ferrari}, L.~{Gualtieri}, and P.~{Pani},
\newblock \PRD {\bf~88}, 023007 (2013).

\bibitem{Doneva_2014}
D.~D. {Doneva}, S.~S. {Yazadjiev}, N.~{Stergioulas}, and K.~D. {Kokkotas},
\newblock \APJL {\bf~781}, L6 (2014).

\bibitem{Haskell_2014}
B.~{Haskell}, R.~{Ciolfi}, F.~{Pannarale}, and L.~{Rezzolla},
\newblock \MNRAS {\bf~438}, L71 (2014).

\bibitem{Chakrabarti_2014}
S.~{Chakrabarti}, T.~{Delsate}, N.~{G{\"u}rlebeck}, and J.~{Steinhoff},
\newblock \PRL {\bf~112}, 201102 (2014).

\bibitem{Pappas_2014}
G.~{Pappas}, T.A.~{Apostolatos}
\newblock \PRL {\bf 112}, 121101 (2014).

\bibitem{Stephani_2003}
H.~{Stephani}, D.~{Kramer}, M.~{MacCallum}, C.~{Hoenselaers}, and E.~{Herlt},
\newblock {\em {Exact Solutions of Einstein's Field Equations}} (Cambridge
  University Press, Cambridge, 2003).

\bibitem{Bardeen_1973}
J.~M. {Bardeen},
\newblock {Rapidly Rotating Stars, Disks, and Black Holes.},
\newblock in {\em Black Holes (Les Astres Occlus)}, edited by C.~{Dewitt} and
  B.~S. {Dewitt}, pp. 241, Gordon and Breach Science Publishers, New York (1973).

\bibitem{Carter_1973}
B.~{Carter},
\newblock {Black hole equilibrium states},
\newblock in {\em Black Holes (Les Astres Occlus)}, edited by C.~{Dewitt} and
  B.~S. {Dewitt}, pp. 57, Gordon and Breach Science Publishers, New York (1973).

\bibitem{Gurlebeck_2013}
N.~G\"urlebeck,
\newblock \prd {\bf~90}, 024041 (2014).

\bibitem{Fodor_1989}
G.~{Fodor}, C.~{Hoenselaers}, and Z.~{Perj{\'e}s},
\newblock \JMP {\bf~30}, 2252 (1989).

\end{thebibliography}

\end{document}